\documentclass[twocolumn,showpacs,preprintnumbers,amsmath,amssymb]{revtex4}
\usepackage{tabularx,graphicx}

\usepackage{color}
\usepackage{hyperref}
\hypersetup{
    colorlinks=true,
    linkcolor=blue,
    filecolor=blue,      
    urlcolor=blue,
}

\begin{document}

\newcommand{\beq}{\begin{equation}}
\newcommand{\eeq}{\end{equation}}
\newcommand{\beqn}{\begin{eqnarray}}
\newcommand{\eeqn}{\end{eqnarray}}
\newcommand{\bmath}{\begin{subequations}}
\newcommand{\emath}{\end{subequations}}
\newcommand{\bra}[1]{\langle #1|}
\newcommand{\ket}[1]{|#1\rangle}

\title{Defying inertia: how rotating superconductors generate magnetic fields}
\author{J. E. Hirsch }
\address{Department of Physics, University of California, San Diego,
La Jolla, CA 92093-0319}

\begin{abstract} 
I discuss the process of magnetic field generation in rotating superconductors in simply connected and multiply connected
geometries. In cooling  a normal metal into the superconducting state while it is rotating,  electrons slow down or speed up depending on the geometry and their location in the sample, apparently defying inertia. I argue that the conventional theory of superconductivity does not explain these processes. Instead, the
theory of hole superconductivity does. Its predictions agree with experimental observations of 
Hendricks, King and Rohrschach for solid and hollow cylinders.
 \end{abstract}
\maketitle 
\section{Introduction}

 A superconductor rotating with angular velocity $\vec{\omega}$ has a uniform magnetic field throughout its interior 
 given by \cite{londonbook,hild,brick,lm2,lm3,lm4,lm5,lm6,lm7}
 \beq
\vec{B}_0=-\frac{2m_ e c}{e}\vec{\omega}
\eeq
where $m_e$ is the bare electron mass and $e$ the electron charge, with its sign. The magnetic field Eq. (1)  is called
the `London field'. Quantitatively, $B_0=1.137\times 10^{-7} \omega$, with $B_0$ in Gauss and $\omega$ in rad/s. 
In the rotating reference frame, the value of $\vec{B}_0$ given by Eq. (1) gives rise to a magnetic Lorentz force $\vec{F}_B$ on
electrons moving with velocity $\vec{v}$ that exactly cancels the Coriolis force $\vec{F}_C$:
\beq
\vec{F}_B+\vec{F}_C=\frac{e}{c}\vec{v}\times\vec{B}_0+2m_e \vec{v}\times\vec{\omega}=0 .
\eeq

It is generally assumed that the London field is equivalent to the Meissner effect, i.e. that one can be deduced from the other.
This is not quite so. The Meissner effect says nothing about the $sign$ of the charge of the mobile charge carriers in
superconductors. Instead, the fact that the magnetic field $\vec{B}_0$ is parallel and not antiparallel to the angular velocity
$\vec{\omega}$ tells us that the mobile charge carriers are negatively charged \cite{eha}. It is true that both the Meissner effect and the
London moment can be deduced from the London equation
\beq
\vec{\nabla}\times\vec{v}_s=-\frac{e}{m_ec}\vec{B}
\eeq
together with Ampere's law
\beq
\vec{\nabla}\times\vec{B}=\frac{4\pi}{c}\vec{J}
\eeq
with 
\beq \vec{J}=n_se(\vec{v}_s-\vec{v}_0)\eeq the supercurrent density, $n_s$ the superfluid density, $\vec{v}_s$ the superfluid velocity
and $\vec{v}_0=\vec{\omega}\times\vec{r}$ the velocity of the body at point $\vec{r}$. In the interior
of a rotating superconductor the superfluid rotates together with the body, hence $\vec{v}_s=\vec{v}_0$ 
 and substituting in Eq. (3), Eq. (1) results for the magnetic field in the interior. The current giving rise to this magnetic field flows within the
London penetration depth of the outer surface $\lambda_L$, given by
\beq
\frac{1}{\lambda_L^2}=\frac{4\pi n_se^2}{m_e c^2} .
\eeq
as follows from Eqs. (3) - (5). The magnetic field decays to zero at the surface for a long cylinder, and is zero in the exterior.

More generally, it is assumed that the following equation holds \cite{geurst}
\beq
\hbar \vec{\nabla}{\varphi}=2m^*(\vec{v}_s-\vec{v}_0) +2m_e\vec{v}_0+\frac{2e}{c}\vec{A}  .
\eeq
for the phase $\varphi$ of the macroscopic wavefunction describing the superfluid, where $m^*$ is the effective mass of
superfluid carriers and $\vec{A}$ the magnetic vector potential. Taking the curl of Eq. (7),
\beq
\vec{\nabla}\times(\vec{v}_s-\vec{v}_0)=-\frac{e}{m^*c}(\vec{B}-\vec{B}_0)
\eeq
where $\vec{B}_0$ is given by Eq. (1), or equivalently
\beq
\vec{\nabla}\times \vec{J}=-\frac{n_s e^2}{m^*c}(\vec{B}-\vec{B}_0)=-\frac{c}{4\pi \lambda_L^2}(\vec{B}-\vec{B}_0)
\eeq
where $\lambda_L$ is given by Eq. (6) with $m^*$ replacing $m_e$. 
In the interior the superfluid rotates with the body, hence $\vec{J}=0$ and $\vec{B}=\vec{B}_0$.

The issue of whether it is the effective mass $m^*$ or the bare mass that determines the penetration depth
$\lambda_L$ has been discussed by us elsewhere \cite{mstar}. It is not important for the purposes of this
paper, so we will assume $m^*=m_e$ in what follows for simplicity, in which case the phase equation (7) is simply
\beq
\hbar \vec{\nabla}{\varphi}=2m_e\vec{v}_s  +\frac{2e}{c}\vec{A}  .
\eeq
 $\vec{v}_s$ is the superfluid velocity in
an inertial (i.e. non-rotating) reference frame. Eq. (10) is valid whether the superconducting body is rotating or at rest in
this inertial reference frame.

Equation (10) does not uniquely determine the state of the superconductor. Integrating the left side of Eq. (10) 
yields
\beq 
\hbar \oint \vec{\nabla}\varphi \cdot \vec{dl}=nh
\eeq
where $n$ is an integer. Which value of $n$ will be attained depends on the geometry of the situation and the process.

          \begin{figure} [t]
 \resizebox{8.5cm}{!}{\includegraphics[width=6cm]{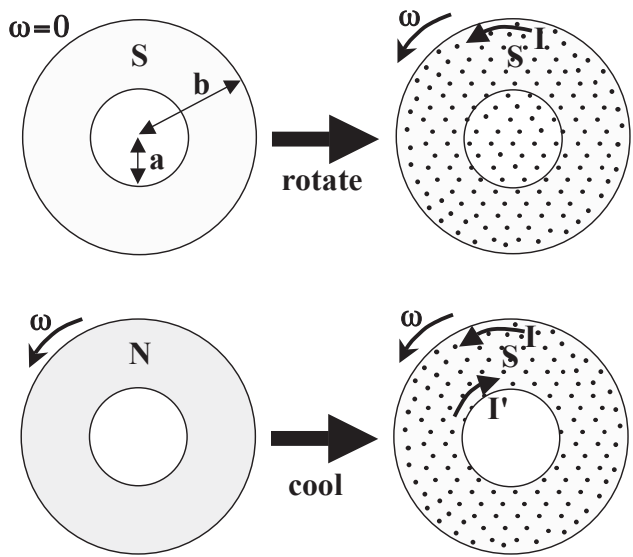}} 
 \caption { Top panel: a superconducting hollow cylinder of inner radius $a$ and outer radius $b$, when set into rotation, develops a uniform magnetic field given by Eq. (1)  (black dots)
 both in the interior of the material and in the hollow space inside. Bottom panel:  the same hollow cylinder, when set into rotation
 while in the normal state and then cooled into the superconducting state, develops a magnetic field only in the  region of the material. In the top right panel, current  $I$ circulates only near the outer surface. In the bottom right panel, currents $I$ and $I'=-I$ in opposite directions circulate
 near the outer and inner
 surfaces respectively. }
 \label{figure1}
 \end{figure}

 Consider a cylindrical shell of inner radius $a$ and outer radius $b$ and  the two processes shown in Fig. 1 top and bottom panels. In the top panel, 
 the system is initially in the superconducting state at rest and is set into rotation, in the bottom panel the system is
 rotating while in the normal state and  is cooled into the superconducting state while rotating. The resulting magnetic field (black dots) is very different.
In the first case, it permeates both the material as well as the interior hollow core, in the second case only the material. This is the behavior
found in the experiments by Hendricks, King and Rohrschach \cite{hkr}, discussed in more detail in a later section. It is also
the behavior that follows from the theory of hole superconductivity \cite{hsc}, as we will explain. Instead, we will argue that the 
conventional theory of superconductivity cannot 	 predict the two different behaviors shown in Fig. 1.

\section{the puzzle}

The conventional theory of superconductivity does not describe the dynamics of the process by which
a magnetic field is generated in a rotating superconductor, just as it does not describe the 
dynamics of the Meissner effect \cite{ondyn}.
It only deals with energetics.

Assuming $b-a>>\lambda_L$, we consider an integration contour in the interior of the shell, where $\vec{v}_s=\vec{v}_0$.
Integration of Eq.  (10) along that contour yields for the situation in the top panel of Fig. 1
\beq
n=0
\eeq
and for the situation in the bottom panel of Fig.1
\beq
n\hbar=2m_e a^2\omega .
\eeq
For example, if $\omega=100 rads/sec$ and $a=1cm$, Eq. (13) yields $n=172.5$. This is of course not possible
because $n$ has to be an integer, so the inner current $I'$ in Fig. 1 will slightly adjust itself to make $n$ 
either $172$ or $173$, and there will be in general a tiny magnetic field in the hollow core. Its maximum value
is such that the flux through the hollow core is half the flux quantum $\phi_0=hc/2e$.

The question we wish to ask is, how does the system decide which of the two final states shown in Fig. 1 
and described by Eqs. (12) and (13) to attain,
and how does it get there?

The energy density of a magnetic field is $B^2/(8\pi)$. Clearly, the magnetic energy is higher in the top right panel of
Fig. 1 than in the bottom right panel. There is a kinetic energy cost in generating the current $I'$ in the bottom panel,
however, that cost is proportional to $(\lambda _L a)$, while the lowering of energy obtained in suppressing
the magnetic field in the hollow core is proportional to $a^2$, with essentially the same prefactor, and we
assume $a>>\lambda_L$.
Why then, if the system is in the state shown in the right top panel of Fig. 1, doesn't the current $I'$ spontaneously
start to flow to lower the total energy of the system? Or, why don't both currents, $I$ and $I'$, appear together and
increase in magnitude together as we start rotating the superconductor from its initial state at rest to the
final angular velocity $\omega$ to reach the final state shown on the bottom right of Fig. 1 rather than
the one on the top right?

 The answer is, because energetics alone does not determine the final state of the system. There has to be a dynamical
 process leading from the initial to the  final state, described by equations that respect the laws of physics.
 
 By the same token, it is not valid to argue that in cooling a rotating normal metal
  (lower left panel in Fig. 1)
   into the superconducting state
 while rotating, the final state in the lower right panel of Fig. 1 will be attained with $n$ given by Eq. (13) rather
 than the state $n=0$ or any other $n$, using as argument that the state described by Eq. (13) has lower total energy.
 There has to be a dynamical process that leads from the bottom left panel of Fig. 1 to the bottom right panel.
 A dynamical process that explains how electrons rotating with the body in the normal state will defy inertia and spontaneously
 slow down if they are near the outer surface
 of the cylindrical shell and will spontaneously speed up if they are near the inner surface during the transition from the normal
 to the superconducting state, to give rise to the currents $I$ and $I'$. The conventional theory gives no indication of what this dynamical process is.

\section{what needs to be explained}

Fig. 2 shows qualitatively what happens when the rotating normal metal is cooled into the superconducting state.
In the normal state there are no currents, electrons and ions rotate together at the same speed $\omega r$ at 
radius $r$ from the center. 
As the system becomes superconducting, 
electrons within $\lambda_L$ of the outer surface slow down relative to the body, and electrons
within $\lambda_L$ of the inner  surface speed up relative to the body by the same amount, thus generating
a magnetic field within the shell but not in the hollow core.

 Let us consider the magnitude of these velocity changes.  From Ampere's law, the current that gives rise to magnetic field
 $B_0$ for a long cylinder of height $\ell$ is $I=c/(4\pi)B_0 \ell$. $I=J\ell \lambda_L$, with $J=en_s\Delta v_s$ the current
 density and $\Delta v_s$ the relative velocity of the carriers with respect to the body. From this and using Eq. (6) it follows that
  the magnitude of  the superfluid velocity near the outer
surface of radius $b$  is  
\beq
v_s^{outer}=\omega b-2\lambda_L \omega
\eeq
and  to nullify the magnetic field generated by Eq. (14) in the hollow core of the cylinder,  electrons within $\lambda_L$ of the inner surface of radius $a$ have to have speed
\beq
v_s^{inner}=\omega a+2\lambda_L \omega .
\eeq
In other words, the electrons near the
outer surface slow down and those near the inner surface speed up during the transition by the same amount.
Instead, when the system already in the superconducting state is rotated from rest, the electrons near the outer
surface slow down but the electrons near the inner surface do not change their speed.

Therefore, there is something about the $process$ of the normal metal becoming superconducting that makes
electrons near the inner surface speed up. This `something' does not take place if the system is already
superconducting when it starts to rotate.

In contrast, electrons near the outer surface will slow down both when we start rotating the superconductor
and when we cool the rotating normal metal. But this doesn't imply that the physics is the same in both
cases. In fact, it is not.

To discuss the behavior of electrons near the outer surface it is simpler to consider the case of a solid cylinder,
i.e. assume $a=0$. 
The process by which electrons near the (outer) surface acquire a slightly lower speed than the body when the superconductor
is rotated starting from rest was explained by Becker et al \cite{becker} already in 1933, even before the Meissner effect was
discovered. It follows simply from Maxwell's equations and from assuming that superconducting electrons
are completely detached from the body. As the body starts to rotate
a time-dependent magnetic field is generated that generates a Faraday electric field that pushes the superfluid electrons
in the same direction as the ions, and they attain the same velocity as the ions in the interior of the cylinder but lag slightly
behind in the surface layer giving rise to the magnetic field. 

In contrast, the process by which electrons near the (outer) surface slow down when the rotating normal metal is 
cooled into the superconducting state while rotating is certainly $not$ explained by Maxwell's equations.
Here, no time-dependent magnetic field results from motion of the ions because electrons are moving together
with the ions initially. So Maxwell's equations alone predict that no magnetic field is generated.  Even more,
Maxwell's equations predict that if electrons near the outer surface were to slow down relative to the body
due to some other unknown reason, a magnetic field in the interior and a Faraday electric field would be generated (counter-emf)  pushing the electrons not to
do that, i.e. pushing the electrons to keep moving together with the ions.

The conventional theory of superconductivity does not explain what that other unknown reason is that acts during the transition
process, and how it overcomes
the Faraday field. The theory
of hole superconductivity does \cite{emf,emf2,inertia}. We review the explanation in  a later section, and then  we show that the same physics also explains why electrons near the inner surface speed up
as the transition occurs in the rotating hollow shell.

          \begin{figure} [t]
 \resizebox{8.5cm}{!}{\includegraphics[width=6cm]{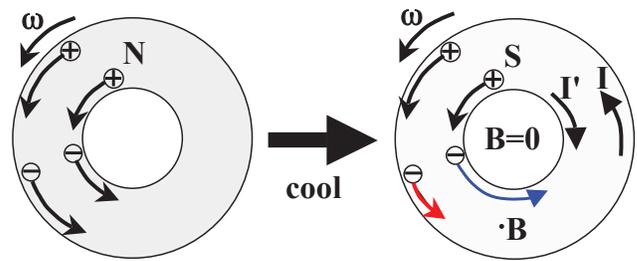}} 
 \caption { When a rotating cylindrical shell in the normal state is cooled into the superconducting state,
 electrons near the outer surface slow down and electrons near the inner surface speed up by $2\lambda_L \omega$ relative to the positive ions,
 as shown by the red and blue  arrows, giving rise to the currents $I$ and $I'=-I$.}
 \label{figure1}
 \end{figure} 
 
\section{superconductor set into rotation}

 To set the stage, let us briefly review why electrons near the surface of a simply connected superconductor end up 
 moving slightly slower than the body when a superconductor
 is set into rotation \cite{becker}. 
 We assume that the superfluid electrons obey the equation of motion
\beq
m_e\frac{d\vec{v}_s}{dt}=e\vec{E}+\vec{F}_{latt}
\eeq
where the first term is the force on the electrons from the induced azimuthal electric field $E$, and the second term is a direct
force that may be exerted by the ions on the superfluid electrons. The electric field $E$ is determined by
Faraday's law 
\beq
\oint \vec{E}\cdot \vec{dl}=-\frac{1}{c}\frac{\partial \phi}{\partial t}
\eeq
and assuming cylindrical symmetry we have
\beq
E(r,t)=-\frac{1}{2\pi r} \frac{1}{c}\frac{\partial \phi (r,t)}{\partial t}
\eeq
where $E(r,t)$ is the induced azimuthal Faraday  electric field and $\phi(r,t)$ the magnetic flux through $r'<r$, at time $t$. 
At radial position r Eq. (18) yields, assuming a uniform magnetic field for all $r'<r$,
\beq
E(r,t)=-\frac{r}{2c} \frac{dB(t)}{dt} .
\eeq
In the interior of the body, superfluid electrons rotate together with the body, i.e. with azimuthal velocity $v_s=\omega r$. 
Using this and combining Eqs. (16) and (19)  yields
\beq
m_e r \frac{d \omega}{dt}=-\frac{er}{2c} \frac{dB(t)}{dt}+F_{latt}.
\eeq
Under the assumption that $F_{latt}=0$, Eq. (20) integrates to
\beq
m_e \omega=-\frac{e}{2c}B
\eeq
i.e. Eq. (1), assuming the initial conditions are $\omega=B=0$. 
Therefore, Eq. (1) is obtained for a superconductor set into rotation {\it under the assumption}
that there is no direct force $F_{latt}$ acting between the ions and superfluid electrons, i.e. that superfluid
electrons and the ionic lattice are completely detached from each other. 

For the case of the cylinder with a hollow core exactly the same derivation applies, assuming the same magnetic field also exists in the interior
of the hollow core as shown in the upper panel of Fig. 1. If instead we were to assume that no magnetic field exists
in the interior of the hollow core, as in the lower right panel of Fig. 1, Eq. (18) yields
\beq
E(r,t)=-\frac{r}{2c}(1-\frac{a^2}{r^2}) \frac{dB(t)}{dt} 
\eeq
and combining with Eq. (16)
\beq
m_e  \frac{d \omega}{dt}=-\frac{e}{2c}(1-\frac{a^2}{r^2}) \frac{dB(t)}{dt}+F_{latt}.
\eeq
which obviously will not yield Eq. (1) unless
\beq
\frac{e}{2c}\frac{a^2}{r^2} \frac{dB(t)}{dt}+F_{latt}=0
\eeq
for which there is no physical justification.
Therefore, we conclude that the assumption $F_{latt}=0$ is correct and the magnetic field develops throughout the interior
of the shell as shown in the top right panel of Fig. 1, hence that  no current develops on the inner surface.

\section{rotating simply connected  superconductor cooled into the superconducting state}
For this situation, if we assume that Eq. (16) with $F_{latt}=0$ is valid throughout the transition, we would conclude that
$\vec{v}_s=\vec{\omega}\times\vec{r}$ independent of time, $E=0$ and $B=0$. This would be inconsistent with the 
phase equation (10), that for a simply connected rotating superconductor predicts $\varphi=0$ and hence
\beq
\vec{A}=-\frac{m_e c}{e}\vec{\omega}\times\vec{r} .
\eeq
For this equation to be valid, electrons near the surface need to slow down when 
the rotating system becomes superconducting to generate the magnetic field Eq. (1) to satisfy Eq. (25). 
According to our discussion in the previous section, it is reasonable to assume that $F_{latt}=0$ in this case also.
So {\it how can electrons that are detached from the lattice and are initially rotating at the same speed as the body 
spontaneously slow down?}

          \begin{figure} [t]
 \resizebox{8.5cm}{!}{\includegraphics[width=6cm]{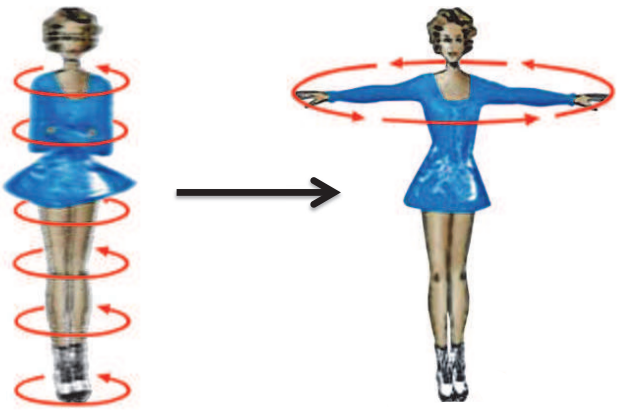}} 
 \caption { A rotating body will slow down if it expands radially outward because its moment of inertia increases.
 Its kinetic energy will decrease in the process. The figure shows schematically normal electrons in the left
 panel and superconducting electrons in the right panel.}
 \label{figure1}
 \end{figure}

The logical simple answer is shown schematically in Fig. 3. If the rotating normal electrons increase their moment of inertia
when they become superconducting, the resulting superconducting condensate will spontaneously slow down to conserve
angular momentum,
hence a magnetic field will be generated since the positive ions continue to rotate at the same speed. 

How this works quantitatively was discussed in ref. \cite{inertia}, we review it briefly here. Assume each electron in going superconducting increases its contribution to the electronic moment of inertia by the amount  
\beq
\Delta i_e=m_e (2\lambda_L)^2 .
\eeq
The total change in the electronic moment of inertia for a cylinder of radius $b$ and height $\ell$ is
\beq
\Delta I_e=m_e (2\lambda_L)^2n_s(\pi b^2 \ell) .
\eeq
On the other hand, if electrons within $\lambda_L$ of the surface slow down by $\Delta v_s$, the reduction in the
electronic angular momentum will be
\beq
\Delta  L_e=-m_e (\Delta v_s) b n_s(2\pi b\lambda_L   \ell)  \omega
\eeq
According to Eq. (14), $\Delta v_s=2\lambda _L \omega$, and replacing in Eq.(28)
\beq
\Delta L_e=-m_e (2\lambda_L)^2n_s(\pi b^2 \ell)  \omega .
\eeq
Thus, assuming (as observed) that electrons in the bulk  continue rotating with angular velocity $\omega$, 
their angular momentum will increase because of the increase in the moment of inertia by
\beq
\Delta L_e^{bulk}=\Delta I_e \omega=-\Delta L_e
\eeq
exactly compensating the decrease in electronic angular momentum due to the slowing down of electrons
in the surface rim of thickness $\lambda_L$ that generates the London field.
The details of this are explained in ref. \cite{inertia}. Note that due to the relation Eq. (6), the change
in moment of inertia $\Delta I_e$ occurs immediately when the system enters the superconducting state
and is temperature independent below $T_c$.

\section{multiply  connected cylindrical superconductor}

In the multiply connected cylindrical shell of inner radius $a$ and outer radius $b$, the increase in the bulk moment of inertia
according to the physics described in the previous section will be
\beq
\Delta I_e=m_e (2\lambda_L)^2n_s(\pi (b^2-a^2) \ell) .
\eeq 
According to Eq. (14), the slowdown of electrons near the outer surface decreases the electronic angular momentum by
\beq
\Delta L_e^{outer}=-m_e (2\lambda_L)^2n_s(\pi b^2 \ell)  \omega 
\eeq
and according to Eq. (15), the speedup of electrons near the inner surface increases the electronic angular momentum by
\beq
\Delta L_e^{inner}=m_e (2\lambda_L)^2n_s(\pi a^2 \ell)  \omega .
\eeq
for a total angular momentum change
\beq
\Delta L_e^{tot}=\Delta L_e^{outer}+\Delta L_e^{inner}+\Delta I_e \omega =0 .
\eeq
Therefore, the increase in speed of electrons in the inner surface is $consistent$ with angular momentum conservation
under the assumption Eq. (26).
If electrons in the inner surface did not speed up in the process of the metal becoming superconducting,
the explanation of ref. \cite{inertia} that the slowdown of electrons near the outer  surface 
 results from increase
in the electronic moment of inertia by the amount Eq. (26) would have been proven wrong.

But consistency is not enough. We still have to explain the physical process by which the electrons in the inner surface
speed up.
Fortunately, the explanation for this speedup near the inner surface is the same as the explanation for the slowdown \cite{inertia}
 near the outer surface, and it is the same as the
explanation for the Meissner effect in the non-rotating superconductor \cite{sm,missing}: {\it $2\lambda_L$ orbits}.

\section{the process}

          \begin{figure} [t]
 \resizebox{6.5cm}{!}{\includegraphics[width=6cm]{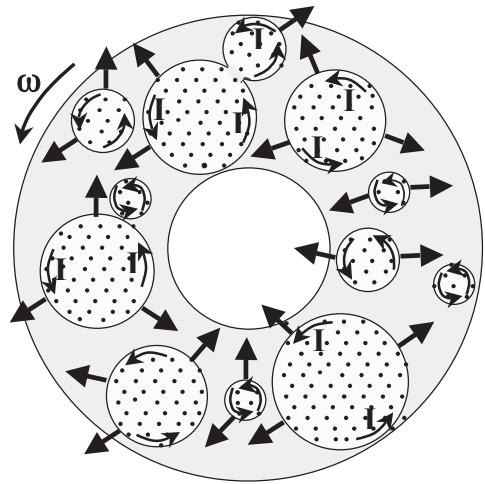}} 
 \caption { Growth of superconducting domains when a rotating normal metal is cooled into the superconducting
 state while rotating. The dots denote magnetic field $B_0$ pointing out of the paper.
 Within each domain, the current near its outer surface is $I=c/(4\pi)B_0 \ell$. }
 \label{figure1}
 \end{figure}

 Fig. 4 shows schematically the development of the superconducting phase in a rotating normal metallic shell.
 The superconducting phase nucleates forming small domains that expand and merge, eventually filling the
 entire region. For simplicity we assume the domains are circular. 
The tangential velocity of carriers
 at the boundary of the domains {\it relative to the body} is $2\lambda_L \omega$, flowing in 
 clockwise direction, opposite direction to the body's rotation.
 The resulting current $I=(c/(4\pi)B_0 \ell$ in the same direction as the body's rotation, shown by the curved arrows, generates the magnetic field $B_0$  in the interior of the domains, denoted by black dots.
 
 It is apparent from the geometry that this process gives a slowdown of (negatively charged) carriers relative to the body in the surface region of each
 domain with  largest $r$ and   a speedup in the surface region of each
 domain with smallest $r$, with $r$ the distance to the cylinder center.  As the domains merge and eventually occupy the entire volume of the cylindrical shell, the internal currents cancel out and only currents near the outer and inner surfaces running parallel to
 the surfaces remain, with the sign as given in the lower right panel of Fig. 1. 
 
           \begin{figure} [t]
 \resizebox{8.5cm}{!}{\includegraphics[width=6cm]{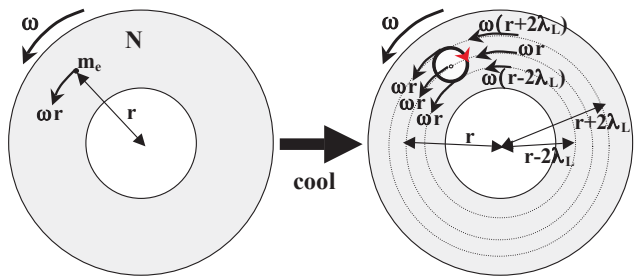}} 
 \caption { Rotating cylinder, the ions at radius $r$ from the origin move with tangential velocity $\omega r$. Assume a point mass $m_e$ at distance $r$ from the center moving together with the ions  at speed $\omega r$ (left panel) expands to a rim of radius $2\lambda_L$ (right panel) keeping its total angular momentum fixed. It acquires clockwise rotation with respect to its center with angular velocity $-\vec{\omega}$,
 indicated by the red arrow.
 The outer and inner points on the rim continue to move at the same speed $\omega r$  they had originally. 
  }
 \label{figure1}
 \end{figure}

 We conclude then that the reason that currents $I$ and $I'=-I$ circulate in the final state along the outer and inner surfaces,
 as opposed to just the current $I$ along the outer surface,  is no longer a mystery. It
 has nothing to do with energetics, it has to do with the fact that 
 there is no dynamical process that will generate the current $I$ without also generating the current $I'$ in the process where
 the rotating normal metal becomes superconducting. The transition occurs through nucleation and expansion of
 domains as shown in Fig. 4, which generates both currents $I$ and $I'$. There is also no dynamical process that will generate the current $I'$ when the system is
 in the state shown in the upper right panel of Fig. 1 with only current $I$ flowing, that would bring the system to the lower energy state
 shown in the lower right panel of Fig. 1 with current $I'$ also flowing, so the process will not happen in a macroscopic system
 even though it would lower the energy.
 
              \begin{figure} [t]
 \resizebox{8.5cm}{!}{\includegraphics[width=6cm]{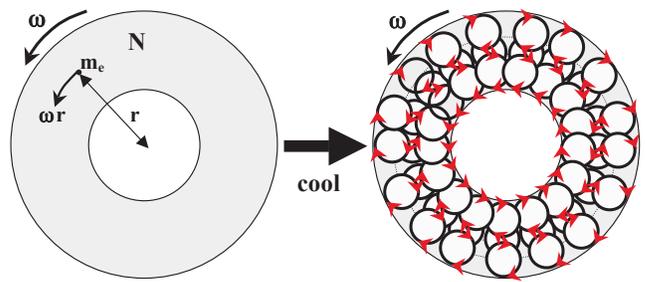}} 
 \caption { Same as Fig. 5, showing the superposition of orbits for many electrons. 
 The internal relative motions with respect to the body cancel out, only the slowdown and speedup near the outer and
 inner surfaces survive.
 }
 \label{figure1}
 \end{figure}

 What remains to be explained is  what causes the speed of the electrons at the boundary of the expanding domains to change by the
 amount $2\lambda_L \omega$ to give rise to the  
 magnetic field $B_0$ in the interior of each domain.

 As mentioned earlier, the slowdown of electrons near the outer surface of a rotating normal metal becoming superconducting
 follows naturally from the assumption that 
 electrons acquire an `intrinsic' moment of inertia 
 \beq
 \Delta i_e=m_e (2\lambda_L)^2
 \eeq
 when they enters the superconducting state. This will happen if the electron `expands' from being a point-like particle
 to a rim of radius $2\lambda_L$, shown schematically in Fig. 5. If the electron is at radius $r$ from the origin, its moment of inertia is 
 \beq
 i_e = m_e r^2
 \eeq when  it is point-like, and it becomes
 \beq
i_e+\Delta i_e=m_e(r^2+(2\lambda_L^2))
 \eeq
 in the superconducting state. We already discussed the angular momentum balance that follows from this
 assumption, which is consistent with the slowdown and speedup required to give rise to the
 observed magnetic field. 
  The dynamics of slowdown and speedup processes is  equally easy to understand.
 When the point-like electron at radius $r$ expands to a   rim of radius $2\lambda_L$ it starts to rotate around the center of the rim in clockwise
 direction with angular velocity $-\vec{\omega}$, driven by the Coriolis force on the outflowing mass. Its total angular momentum is  
 \beq
 l_e = m_e(r^2+(2\lambda_L)^2)\omega -m_e(2\lambda_L)^2\omega=m_er^2\omega .
 \eeq
 In other words, it doesn't change.  As a consequence,
 the outer point of the rim, at distance $r+2\lambda_L$ from the origin, continues to move with tangential velocity $\omega r$, hence
 slower by $2\lambda_L\omega$ than  the ions at that
 radius that are moving with tangential velocity $\omega (r+2\lambda_L)$, and similarly the inner point of the
 rim at radius $r-2\lambda_L$ also continues to move with tangential speed $\omega r$, faster by $2\lambda_L \omega$ than  the ions at that radius that are moving at
 speed $\omega(r-2\lambda_L)$. Superposition of these overlapping orbits of all the electron causes the
 internal velocities to cancel out, so that slowdown and speedup by the amount $\pm 2\lambda_L \omega$ relative to the ions  only survives in  layers of thickness $\lambda_L$ of  the outer
 and inner surfaces. Figure 6 shows schematically the superposition of these motions.

 We assume that this expansion of the electronic wavefunction occurs when the normal metal becomes superconducting,
 whether or not the body is rotating \cite{sm}. Since in the top left panel of Fig. 1 the system is already superconducting before it starts rotating, the process of expansion does not occur when the system is set into rotation, and the inner current cannot
 be generated. The generation of the outer current is simply understood from Maxwell's equations as discussed in an earlier section.

            \begin{figure} [t]
 \resizebox{8.5cm}{!}{\includegraphics[width=6cm]{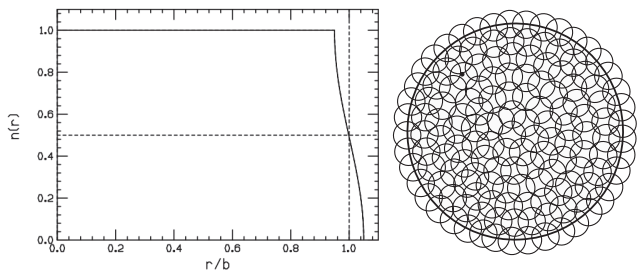}} 
 \caption {  When the point electrons expand to rims of radius $d=2\lambda_L$ the mass distribution
 extends beyond the radius of the cylinder $b$ up to radius $b+d$. The density in the interior 
 does not change except in   a layer of thickness $d$ from the surface where it decreases. The left panel shows the density
 versus $r$, normalized to $1$. $b=1$, $d=0.05$.
 }
 \label{figure1}
 \end{figure}

             \begin{figure} [t]
 \resizebox{8.5cm}{!}{\includegraphics[width=6cm]{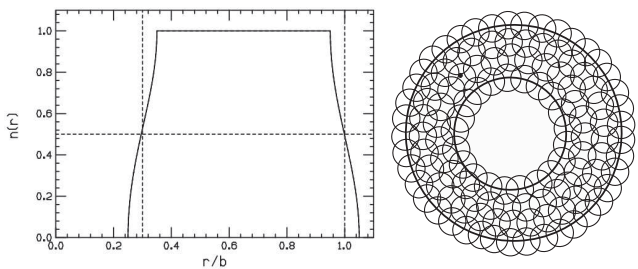}} 
 \caption { Same as Fig. 7 for a cylindrical shell of inner radius $a=0.3$. Here the mass density 
 extends from $a-d$ to $b+d$.
 }
 \label{figure1}
 \end{figure}

\section{mass distribution}
Let us consider the mass distribution that results from the orbit enlargement, assuming the centers of the rims remain 
where the point-like electrons were before expansion. For a solid cylinder, Fig. 7 shows schematically the orbits
and the resulting mass density versus $r$, the distance to the center of the cylinder. The density in the interior is not
changed  except in a layer of thickness $d=2\lambda_L$, the radius of the orbits, where it is reduced, and
the density extends beyond the radius $b$ up to radius $b+d$.
In the region $b-d \leq r \leq b+d$ the density (normalized to $1$ in the normal state)  is given by
\beq
n(r)=1-  \frac{1}{\pi} cos^{-1} \frac{b^2-r^2-d^2} {2rd} .
\eeq
Similarly, in a cylindrical shell of inner radius $a$ the mass density will extend inside radius $a$ to radius
$a-d$ and is smaller in the layer $a<r<a+d$ than in the bulk, as shown in Fig. 8. 
In the region, $a-d \leq r \leq a+d$ it is given by
\beq
n(r)=1-  \frac{1}{\pi} cos^{-1} \frac{r^2+d^2-a^2} {2rd} .
\eeq
The moment of inertia changes as discussed in the previous sections, Eq. (27) and Eq. (31), result from these mass
distributions. The total moment of inertia per unit mass  for a shell of mass $M$ with  mass distribution as in Fig. 8 is
\beq
\frac{I}{M}=\frac{1}{2}(b^2+a^2)+d^2 .
\eeq

If electrons in superconductors had only mass and no charge, this would be the whole story.
The mass distribution would be as shown in Figs. 7 and 8 and electrons near the surfaces would
flow with the speeds given by Eqs. (14) and (15), the slowdown and speedup with respect to the body
explained by the change
in moment of inertia and conservation of electronic angular momentum explained earlier.
If we could now give the electrons back their charge and turn on half of Maxwell's equations,
allowing for Ampere's law but not for Gauss's law, the magnetic fields shown in the lower right
panel of Fig. 1 would result from this speed distribution.
The reality is slightly more
complicated. Because of Gauss' law, the final mass configuration cannot be 
as shown in Figs. 7 and 8 because spillout of charge beyond the sample surface a distance 
$2\lambda_L$ costs enourmous electrostatic energy \cite{missing}. Therefore,
the charge has to be pulled back into the body by electrostatic forces.  
The result of this will be discussed in the next section.

\section{backflow}

Because of electrostatics, the final mass and associated charge distribution cannot extend beyond the surface of the body by several $100 \AA$ as Figs. 7 and 8 show.
Electrons will be pulled back into the body by electrostatic attraction. Let us just consider the solid cylinder (Fig. 7) for simplicity, the extension to the
cylindrical shell is straightforward.   Electrons moving
inward decrease their moment of inertia with respect to the axis of the cylinder ($m_e r^2$). Instead of speeding up keeping their
angular momentum, they keep their angular
velocity constant by transferring angular momentum to the body that will  slightly speed up.

            \begin{figure} [t]
 \resizebox{8.5cm}{!}{\includegraphics[width=6cm]{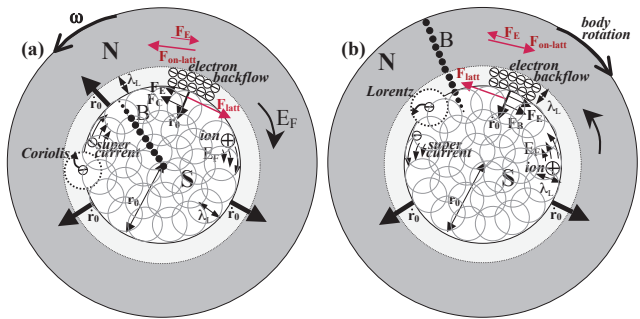}} 
 \caption {(a) Rotating normal metal becoming superconducting and (b) Meissner effect.
The phase boundary at radius $r_0$ expands outward. Black
dots indicate magnetic field pointing out of the paper. Electrons at the phase boundary expand their
orbits to radius $2\lambda_L$  and are (a) slowed down relative to the body by the Coriolis force or (b) accelerated by the magnetic Lorentz force. A  backflow of normal electrons takes
place to compensate for the radial mass and charge imbalance, in the process transferring momentum from electrons to the body.
 See text and ref. \cite{momentum}  for more details.}
 \label{figure1}
 \end{figure} 

The way this happens is entirely parallel to what happens in the Meissner effect, that was discussed in detail in  Ref. \cite{momentum}, as shown in Fig. 9.
For simplicity we can assume that as the metal is cooled the superconducting region forms first at the center of the cylinder and expands maintaining
cylindrical symmetry.  

For both cases, there is a backflow of normal carriers flowing radially inward, with no azimuthal component (relative to the moving body in (a)). It is essential 
that these carriers are holes, i.e. electrons with negative effective mass \cite{momentum}.  For the rotating case,
electric ($F_E$)  and Coriolis ($F_C$)  forces of equal sign and magnitude act on the backflowing electrons  in counterclockwise direction, for  the Meissner case
electric   ($F_E$)  and magnetic  Lorentz  ($F_B$)   forces  of equal sign and magnitude  act on the backflowing electrons  in clockwise direction. There is no magnetic force in the rotating case because there is
no magnetic field in that region. In both cases these forces are balanced by   a  force $F_{latt}$  exerted by the lattice on the negative effective mass electrons
 so that the backflow is radial, with no
azimuthal component, and the transfer of momentum occurs reversibly.  By Newton's third law, a force $F_{on-latt}=-F_{latt}$ transfers momentum from the backflowing electrons to the body. 
Half of that momentum transfer is cancelled by the Faraday electric force $F_E$ acting on the ions  in opposite direction. The  result in the Meissner case,
as discussed in detail in Ref. \cite{momentum}, is to transfer to the body  the growing angular momentum of the supercurrent with opposite sign.
Similarly for the rotating superconductor, the result is to transfer the growing deficit of electronic angular momentum due to the slowdown of electrons in
the layer of thickness $\lambda_L$ adjacent to the phase boundary to the body, that will slightly speed up. The end result is a  uniform electronic
mass and charge distribution with angular momentum  deficit $\Delta L_e$ given by Eq. (29) and a corresponding increase in the body's angular momentum
through a slight increase in its angular velocity
\beq
\Delta \omega =\frac{m_e}{m_pA} \frac{n_s}{n} \frac{8\lambda_L^2}{b^2 - a^2}\omega
\eeq
with $m_p$ the nucleon mass, A the atomic weight, $n$ the ionic number density and $b$ and $a$ the 
outer and inner radii of the cylindrical shell ($a=0$ for the solid cylinder).

\section{the experiments}
The experiments of Hendricks, King and Roschach \cite{hkr} are entirely consistent with the behaviour shown in Fig. 1.
They conducted experiments on a thin cylindrical shell and a solid cylinder. For the solid cylinder, which was a 
high purity single crystal, they found that a magnetic field is generated throughout the sample both when they
cooled while the cylinder was rotating and when they rotated the superconductor at rest (lower and upper panels of
Fig. 1, with $a=0$). For the shell they found that a magnetic field in the hollow core was generated when the sample
was first cooled to become superconducting and then rotated, consistent with Fig. 1 upper panel.
When they cooled the rotating shell they found essentially no effect, and if subsequently the rotation was stopped
a magnetic field in opposite direction was generated, of the same magnitude as when the sample was first cooled and
then rotated. This effect corresponds to stopping the rotation of the rotating superconducting shell
shown in the right lower panel of Fig. 1, which will cause the outer current to stop but not affect the inner
current. This is shown schematically in Fig. 10.

            \begin{figure} [t]
 \resizebox{8.5cm}{!}{\includegraphics[width=6cm]{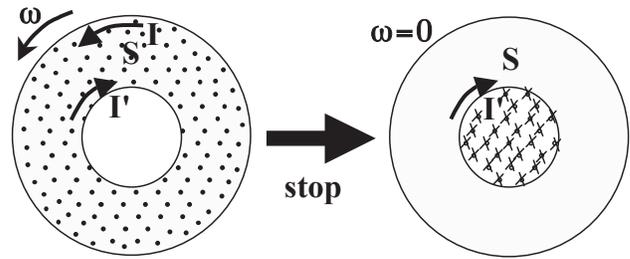}} 
 \caption { When the rotating cylindrical shell of the lower right panel of Fig. 1 is stopped, the outer
 current stops but the inner current remains, giving rise to a magnetic field in the hollow core pointing into the 
 paper (crosses).
 }
 \label{figure1}
 \end{figure}

It is interesting to consider what will happen upon cooling a rotating solid cylinder that is a type II superconductor.
To our knowledge that experiment has not been done. Clearly, in cooling it would first enter the 
mixed-phase region $H_{c1}<B_0<H_{c2}$. Hence an Abrikosov lattice will appear where the
`vortices' have no magnetic field in their interior, and the superconducting regions have magnetic field
$B_0$. Figure 11 shows the pattern of currents that would develop. 
It is difficult to imagine any other mechanism for electrons that are rotating with the body in the normal 
state to defy inertia and develop the   intricate pattern of currents shown in Fig. 11  upon entering
the superconducting state 
{\it other than} the orbit expansion with angular momentum conservation that we have discussed in
this paper.

\section{discussion}

 Rotating superconductors know the difference between positive and negative charge,  the Meissner effect does not.
 If electrons had positive charge and ions negative charge the Meissner effect would look exactly the same, but rotating 
 superconductors would generate magnetic fields in direction antiparallel rather than parallel to their angular velocity. Experimentally, that is never the case. Thus, rotating superconductor tell us that charge asymmetry is an essential aspect of 
 superconductivity, as proposed by the theory of hole superconductivity \cite{chargeasym} and contrary to 
 BCS theory that is electron-hole symmetric.
 For that reason it is important to understand  rotating superconductors in addition to understanding the Meissner effect, they are not equivalent. In fact, we find that understanding rotating superconductors is even simpler than understanding the Meissner effect.
 
             \begin{figure} [t]
 \resizebox{8.5cm}{!}{\includegraphics[width=6cm]{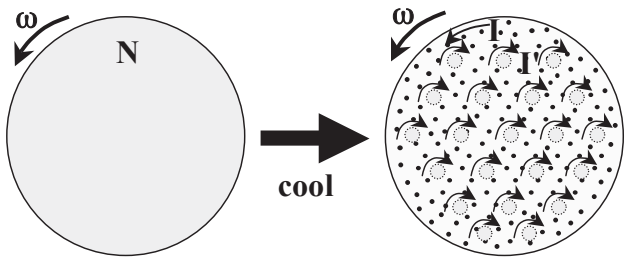}} 
 \caption { `Vortex' pattern that will  develop when a type II  solid cylinder rotating with angular frequency $\omega$ is cooled
 (in the absence of applied magnetic field). It will first enter into the 
 mixed superconducting 
 phase shown.  In the superconducting region (lighter grey) the magnetic 
 field is $B_0=-(2m_ec/e)\omega$ pointing out of the paper, indicated by the black dots. The `vortices'  (darker grey circles forming an 
 Abrikosov lattice) have zero magnetic field in their normal interior, the `non-flux' inside the
 vortex has magnitude $\phi_0=hc/2e\sim 4\lambda_L^2 B_0$.  This pattern of currents
 (counterclockwise $I$ near the surface, clockwise $I'=-I$ around the vortices) results from orbit expansion in the 
 superconducting region as discussed earlier. On further cooling, orbits in the vortex cores will also expand canceling
 the currents around them, and eventually the entire cylinder will enter the Meissner phase 
 with $B_0$ permeating the entire volume and no interior currents.  If the rotation is stopped
 when the system is in the state shown in the figure, the current along the outer surface will stop
 and the vortices will acquire magnetic field $B_0$ in their interior pointing into the paper.}
 \label{figure1}
 \end{figure} 
 
 In a nutshell: let us take as an experimental fact that when a normal metallic shell rotating with angular
 frequency $\omega$ becomes superconducting,
 electrons within a distance $d$ from the outer surface spontaneously slow down by $2\omega d$,
and electrons within a distance $d$ from the inner surface spontaneously speed up by $2\omega d$. 
Experimentally we know this is so  from measuring a magnetic field, but other than that we don't need to know anything
about electromagnetism, mechanics suffices. How can this observation
be understood in the simplest possible way?

The answer proposed in this paper is that in the transition to superconductivity 
electrons expand from being  point particles of mass $m_e$ to 
being circular  rims of radius $2d$ and mass $m_e$. This increases their moment of inertia by $m_e(2d)^2$, and conservation of
angular momentum does the rest, as discussed in earlier sections. That is how electrons defy inertia
(in appearence) and no longer move at the same speed as the rotating body upon entering the superconducting state.

We argue that this explanation is the simplest and most natural way to explain this counterintuitive \cite{absurd}  experimental fact, and for that
reason it is compelling and  likely to be correct. In contrast,  the conventional theory of superconductivity
provides $no$ explanation for what is the physics that gives rise to the slowdown and speedup nor how angular momentum
is conserved, just as it does not do it for the Meissner effect \cite{momentum,entropy}.   
 
  And of course the other reason we find this explanation compelling is that we used the same physics
  to explain the Meissner effect in earlier work \cite{sm,missing,emf,emf2,ondyn,momentum,kinetic,chargeasym,entropy}. Expansion of orbits from a microscopic radius to radius $2\lambda_L$ in the presence
  of a magnetic field $B$ generates through the magnetic Lorentz force an azimuthal current and a magnetic field $-B$ that cancels the magnetic field in the interior, as a consequence the
  magnetic field is expelled. We have explained in detail the dynamics of this process for a simply connected
  superconductor  in earlier work. For a metallic shell in a magnetic field it works similarly as for the rotating
  shell discussed in this paper:  expanding orbits in a magnetic field acquire azimuthal velocity due to the action of the magnetic Lorentz force and
 give rise to currents in opposite directions along the outer and inner surfaces to cancel the magnetic field in the
 interior of the shell while leaving the magnetic field  (almost) unchanged in the inner hollow core.

There are small differences  between  rotating superconductors and  the Meissner effect  in the details of how conservation of angular momentum works.  For rotating superconductors   the electronic angular momentum does not change upon orbit expansion.  Instead, in the Meissner effect it certainly does change, since there is no electronic angular momentum
in the normal state and there is angular momentum  in the Meissner state that carries current.
We have explained in earlier work how angular momentum conservation works for the Meissner effect \cite{momentum} and 
its reverse, superconductor-normal transition in a field \cite{disapp}:   the growing or decreasing angular momentum of the supercurrent is
compensated by angular momentum of the body in opposite direction. The way this happens is,  momentum is stored in the electromagnetic field as orbits expand or contract, and is picked up from there by backflowing normal electrons
that transfer it to the body. We showed that for this to happen in a reversible way, as required by thermodynamics \cite{entropy},
$requires$ that the normal backflowing carriers are $holes$. 

For the case of rotating superconductors  no momentum is stored in the electromagnetic field
because the magnetic field and the radial electric field resulting from orbit expansion or contraction are in different spatial regions \cite{inertia}.
There is backflow of normal electrons, they themselves lower their angular momentum in this process (while keeping their angular velocity constant)
 and transfer it to the massive body that very slightly increases
its speed of rotation, thus compensating for the slowdown of superfluid electrons in the surface layers.  Just as for the Meissner effect it is indispensable that the backflowing
electrons  are hole-like \cite{revers}.

  The orbit expansion is 
driven by lowering of quantum kinetic energy \cite{sm,missing,kinetic} and occurs whether or not the body is rotating
and whether or not there is a magnetic field. In the absence of rotation and magnetic field it results in the
generation of a pure spin current that flows within $\lambda_L$ of the surfaces \cite{electrospin}. The angular momentum in these
$2\lambda_L$ orbits is found to be exactly $\hbar/2$  using the spin-orbit interaction derived from
Dirac's equation  \cite{sm,bohr}. 

In previous work we  found that the orbit expansion is associated with expulsion of negative charge
from the interior to the surface of superconductors \cite{chargeexp,sm} and found
a quantitative relation between the magnitude of the charge expelled and the quantized spin current \cite{electrospin}. This would appear to be in contradiction with the results in Sect. VIII that the electronic 
density and hence the charge density in the interior does not change upon  orbit expansion. The reason reality is slightly different 
from what is shown in Sect. VIII is a quantum effect, presumably  associated with
electronic Zitterbewegung \cite{emf2}. The slight reduction of negative charge density in the interior
is found to be $n_s r_q/R$ \cite{electrospin}, with $r_q=\hbar/2m_ec$ the quantum electron radius
and $R$ the radius of the cylinder. For the cylindrical shell, the same equations \cite{electrospin} 
predict it  to be $n_s r_q/(b-a)$ (in the regime $b-a>>\lambda_L$).

There are two slightly different ways to think about the physics we have discussed. One is
that the electron's mass and charge expand to occupy a  rim of radius $2\lambda_L$, with charge and mass uniformly
distributed along the rim. In this picture, the entire rim rotates with angular frequency $-\vec{\omega}$. A second way is that
the point-like electron expands its orbit from microscopic-like to an orbit of radius $2\lambda_L$, and the electron moves
along this orbit with angular frequency $-\vec{\omega}$. They are essentially equivalent, but the second way gives us
a way to interpret  the `phase' of the electron as its angular position along this orbit. Because these orbits
are highly overlapping the angular position of electrons in different orbits has to be correlated in order to avoid collisions.
This provides an intuitive picture of the `phase coherence' that develops when a system goes superconducting and the orbit
expansion occurs. 
In this picture, in the normal state orbits are non-overlapping \cite{missing}  hence phase coherence is not required.

 Our analysis may be relevant beyond the confines of understanding superconductivity. 
 The nature of electrons is as of yet not  understood \cite{einstein}, nor is the structure of space-time. In free space we can think of 
 the electron
 as a  particle of mass $m_e$, charge $e$, radius $r_q=\hbar/(2m_e c)$ and intrinsic angular momentum $\hbar/2$. 
 Remarkably, we find that in superconductors electrons behave as rims of mass $m_e$, charge $e$, radius $2\lambda_L$
 $and$ orbital angular momentum $\hbar /2$ \cite{sm,bohr}. The fact that $both$ the charge and the mass of the electron 
 expand to radius $2\lambda_L$ is clearly revealed by the Meissner effect and the physics of  rotating 
 superconductors respectively. The transition to superconductivity is thus revealed as an expansion of the electron by 
 a factor $(2/\alpha)^2$ \cite{emf2}, with $\alpha$ the fine-structure constant. The fact that expansion, rotation, quantum kinetic energy lowering, spin-orbit interaction
 and $\hbar/2$ angular momentum quantization are intimately coupled in superconductors is likely to  have  implications in other
 realms.


\begin{references}
\bibitem{absurd} This behavior, for a simply connected rotating superconductor, was characterized as ``quite absurd'' by F. London in  ref. \cite{londonbook}, p. 82. One may wonder what level  of absurdity London would have ascribed to the
corresponding behavior in type II superconductors (Fig. 10).
\bibitem{einstein} As Einstein remarked, `it would be sufficient to really understand the electron'.
\end{references}
\end{document}